\documentclass[sigconf]{acmart}

\copyrightyear{2022}
\acmYear{2022}
\setcopyright{rightsretained}
\acmConference[ASE '22]{37th IEEE/ACM International Conference on Automated Software Engineering}{October 10--14, 2022}{Rochester, MI, USA}
\acmBooktitle{37th IEEE/ACM International Conference on Automated Software Engineering (ASE '22), October 10--14, 2022, Rochester, MI, USA}
\acmDOI{10.1145/3551349.3561161}
\acmISBN{978-1-4503-9475-8/22/10}

\usepackage{amsmath}
\usepackage{mathtools}

\usepackage{enumitem}
\usepackage{fancyvrb}
\usepackage{listings}
\usepackage{etoolbox}
\usepackage[plainruled]{algorithm2e}
\usepackage{pifont}
\usepackage{tikz}
\usepackage{soul}

\makeatletter
\let\old@lstKV@SwitchCases\lstKV@SwitchCases
\def\lstKV@SwitchCases#1#2#3{}
\makeatother
\usepackage{lstlinebgrd}
\makeatletter
\let\lstKV@SwitchCases\old@lstKV@SwitchCases

\lst@Key{numbers}{none}{%
    \def\lst@PlaceNumber{\lst@linebgrd}%
    \lstKV@SwitchCases{#1}%
    {none:\\%
     left:\def\lst@PlaceNumber{\llap{\normalfont
                \lst@numberstyle{\thelstnumber}\kern\lst@numbersep}\lst@linebgrd}\\%
     right:\def\lst@PlaceNumber{\rlap{\normalfont
                \kern\linewidth \kern\lst@numbersep
                \lst@numberstyle{\thelstnumber}}\lst@linebgrd}%
    }{\PackageError{Listings}{Numbers #1 unknown}\@ehc}}
\makeatother

\newcommand{\myparagraph}[1]{\vspace{0.6em}\noindent\textbf{#1}.}
\newcommand{\ignore}[1]{}

\usepackage{fancybox}
\definecolor{mycolor}{rgb}{0.122, 0.435, 0.698}
\newcommand{\result}[1]{%
\cornersize{0.2}
\vspace{1mm}
\hspace{-3.5mm}\Ovalbox{\parbox[t]{0.97\columnwidth}{#1}}%
\vspace{1mm}
}
\usepackage{multirow}
\newcommand{\isan}{$\textsc{ReZZan}$\xspace}
\newcommand{\bisan}{$\textsc{ReZZ}_\text{lite}$\xspace}
\newcommand{\asan}{\textsc{ASan}\xspace}
\newcommand{\fuzzan}{\textsc{FuZZan}\xspace}
\newcommand{\cmark}[0]{\ding{51}}

\copyrightyear{2022} 
\acmYear{2022} 
\setcopyright{rightsretained} 
\acmConference[ASE '22]{37th IEEE/ACM International Conference on Automated Software Engineering}{October 10--14, 2022}{Rochester, MI, USA}
\acmBooktitle{37th IEEE/ACM International Conference on Automated Software Engineering (ASE '22), October 10--14, 2022, Rochester, MI, USA}
\acmDOI{10.1145/3551349.3561161}
\acmISBN{978-1-4503-9475-8/22/10}

\begin{document}

\date{}

\title{Efficient Greybox Fuzzing to Detect Memory Errors}

\author{Jinsheng Ba}
\affiliation{\institution{National University of Singapore}
\country{Singapore}}
\email{jinsheng@comp.nus.edu.sg}
\author{Gregory J. Duck}
\affiliation{\institution{National University of Singapore}
\country{Singapore}}
\email{gregory@comp.nus.edu.sg}
\author{Abhik Roychoudhury}
\affiliation{\institution{National University of Singapore}
\country{Singapore}}
\email{abhik@comp.nus.edu.sg}

\begin{abstract}
Greybox fuzzing is a proven and effective testing method for the detection of
security vulnerabilities and other bugs in modern software systems. Greybox fuzzing can also be used in combination with a \emph{sanitizer},
such as AddressSanitizer (ASAN), to further enhance the detection of
certain classes of bugs such as buffer overflow and use-after-free errors.
However, sanitizers also introduce additional performance overheads, and
this can degrade the performance of greybox mode fuzzing---measured in the order of $2.36{\times}$ for fuzzing with ASAN---partially negating
the benefit of using a sanitizer in the first place.
Recent research attributes the extra overhead to program startup/teardown costs that can dominate fork-mode fuzzing.

In this paper, we present a new memory error sanitizer design that is specifically optimized for fork-mode fuzzing.
The basic idea is to mark object boundaries using \emph{randomized tokens} rather than disjoint metadata (as used by traditional sanitizer designs).
All read/write operations are then instrumented to check for the token, and if present, a memory error will be detected.
Since our design does not use a disjoint metadata, it is also very lightweight, meaning that program startup and teardown costs are minimized for the benefit of fork-mode fuzzing.
We implement our design in the form of the \isan tool, and show an improved fuzzing performance overhead of $1.14$-$1.27{\times}$, depending on the configuration.
\end{abstract}



\maketitle

\section{Introduction}

Fuzz testing is a proven method for detecting bugs and security vulnerabilities in real-world software.
Fuzz testing can be seen as a biased random search over the domain of program inputs, with the goal of uncovering inputs that cause the program to crash or hang.
The biased random search may be guided by an objective function, as in the case of \emph{greybox} fuzzing, which aims to maximize code coverage.
Alternatively, the biased random search can be guided by other forms of feedback, such as symbolic formulae, as is the case with \emph{whitebox} fuzzing based on symbolic execution. 
However, because of the scalability challenges in conducting symbolic execution on real-world software, coverage-guided greybox fuzzing is more widely adopted in the practice of security vulnerability detection.
Usually, greybox fuzzing uses instrumentation that is inserted at compile-time, and then random mutations of provided seed inputs are generated and tested over the lifetime of the fuzzing campaign.
If a mutated input is found to traverse new instrumented locations/edges, it is retained and prioritized for further mutation. In this way, via repeated mutation, the fuzzing campaign {\em covers} (a portion of) the program input space, and seeks to find vulnerabilities.
Coverage-guided greybox fuzzing is a well-known technique for finding vulnerabilities today and is embodied by popular tools such as AFL~\cite{afl} and \textsc{LibFuzzer}~\cite{libfuzzer}.

The core aim of greybox fuzzing is to detect vulnerabilities in the target program.
One important class of bug is \emph{memory errors}, which include \emph{spatial memory errors} such as object-bounds errors (including buffer overflows/underflows), and \emph{temporal memory errors} that access an object after it has been \verb+free()+'ed.
Memory errors are a common occurrence in software implemented in \emph{unsafe programming languages}, such as \verb+C+/\verb~C++~, which, for performance reasons, use manual memory management and no bounds checking by default.
Furthermore, experience with the industry has shown that memory errors are a common source of security vulnerabilities~\cite{microsoft-bluehat-talk}. This is because memory errors may grant attackers the ability to change the contents of memory, and this can form the basis of information disclosure and control-flow hijacking attacks.

Importantly, memory errors do not necessarily cause the program to immediately crash, and such ``silent'' memory errors can be difficult to detect during a fuzz campaign.
To address this problem, the target program may be instrumented using a \emph{sanitizer}, which uses a program transformation to insert additional code before each memory operation. The additional code checks for object-bounds and use-after-free errors, and if detected, will immediately abort (crash) the program, essentially making memory errors visible.
Such instrumentation is quite natural to incorporate into greybox fuzzing. 
Memory error sanitizers, such as AddressSanitizer~\cite{asan}, have been implemented as part of the LLVM Compiler Infrastructure Project~\cite{llvm} and can be enabled by passing a suitable switch (\verb+-fsanitize=address+) to the compiler.

However, the combination of fork-mode fuzzing and memory error sanitizers is known to suffer from significant performance overheads.
For example, under our experiments, the combination of AFL+AddressSanitizer runs at a ${\sim}58\%$ reduction of throughput (execs/sec) compared to AFL alone. Recent work~\cite{fuzzan} attributes this reduction in performance to the negative interaction between the sanitizer implementation and the fuzzing process.
Specifically, traditional memory error sanitizer designs use a \emph{disjoint metadata} to track memory state (e.g., is the memory free or out-of-bounds?).
However, maintaining disjoint metadata can negatively impact program startup/teardown costs due to additional overheads, such as an increased number page faults (metadata access) and additional page fault handling costs~\cite{fuzzan}.
This interacts poorly with the fuzzer which must fork a new process for each generated test input.

In this paper, we present a new memory error sanitizer design that is specially optimized for fork-mode greybox fuzzing.
Since the disjoint metadata is the main source of additional page faults and associated overheads, we propose a memory error sanitizer design that eliminates metadata based on the idea of \emph{Randomized Embedded Tokens} (RETs) pioneered by tools such as LBC~\cite{hasabnis2012light} and REST~\cite{sinha18rest}.
The key idea is to track memory state 
by using a special token that is initialized to some predetermined random nonce value.
With a suitable run-time environment, out-of-bounds (redzone) and \verb+free()+'ed memory can be ``poisoned'' by writing the nonce value to these locations.
Next, a program transformation inserts instrumentation that checks all memory operations to see if the nonce value is present, indicating a memory error.
By representing memory state using the memory itself, we eliminate the additional page faults that otherwise would have been generated by accessing the disjoint metadata, meaning that a RET-based design has the potential to improve the overall sanitizer+fuzzing performance.

That said, existing RET-based sanitizers suffer from various limitations, such as coarse-grained memory error detection~\cite{sinha18rest} or the retention of a disjoint metadata in order to resolve \emph{false detections}~\cite{hasabnis2012light}.
Here, a ``false detection'' may occur if the nonce value happens to be generated by chance during normal program execution.
If this occurs, then legitimate memory may be deemed poisoned, resulting in a ``false detection''.
However, for the application of fuzz testing, we argue that false detections can be tolerated provided the occurrence is sufficiently rare and can be eliminated by automatically re-executing the program with a new random nonce value.
In our experiments, no false detection was observed during 19200 hours (${\sim}2.2$ years) of combined CPU time.
Another problem with the basic RET-based design is the memory error detection granularity.
This is an artifact of multi-byte token sizes, where some bounds overflows may never reach a token, resulting in a ``missed detection''.
For this we propose \emph{refined boundary check}, which encodes boundary information directly into the token, allowing for byte-precise overflow detection.
We show that our design can detect the same class of memory error as traditional sanitizers, such as AddressSanitizer.

We have implemented our design in the form of the (\emph{REt+fuZZing +sANitizer}) \isan tool.
We show that \isan is significantly faster under fork-mode fuzzing, running at a $1.27{\times}$ overhead over ``native'' AFL fuzzing without any sanitizer.
In comparison, AddressSanitizer incurs an overhead of $2.36{\times}$. 
We also present a simplified configuration without refined boundary checking, which runs at a 
$1.14{\times}$ overhead.
This configuration exchanges an increased throughput for a modest reduction of error detection capability.

\myparagraph{Contributions}
In summary, our main contributions are:
\begin{itemize}[leftmargin=*,topsep=1ex,itemsep=0pt,partopsep=1ex,parsep=0ex]
    \item[-] We propose a memory error sanitizer design based on the concept of \emph{Randomized Embedded Tokens} (RETs)~\cite{hasabnis2012light, sinha18rest}.
    We show that a RET-based sanitizer design can minimize program startup/teardown costs, and can therefore optimize the combination of sanitizers and fork-mode fuzzing.
    \item[-] We tune our design so that false detections are very rare in practice, meaning that a disjoint metadata (and associated overheads) is not necessary and can be eliminated.
    We also introduce the notion of \emph{refined boundary checking} for byte-precise memory error detection under a RET-based design.
    \item[-] We have implemented our design in the form of the \isan tool.
    We have integrated \isan with popular greybox fuzzers. 
    \item[-] Our evaluation results against 
    contemporary works (AddressSanitizer (ASAN)~\cite{asan} and FuZZan~\cite{fuzzan}) show a clear benefit from our approach, with a modest $1.27{\times}$ overhead with respect to native throughput, compared to $2.36{\times}$ for AddressSanitizer.
\end{itemize}

\noindent
\textbf{Open Source Release}.
The \isan tool is available open source:
\begin{center}
    \textcolor{blue}{\url{https://github.com/bajinsheng/ReZZan}}
\end{center}
An archived artifact~\cite{zenodo} and pre-print~\cite{preprint} are also published online.

\definecolor{keyword}{rgb}{0.40, 0.40, 0}
\definecolor{comment}{rgb}{0, 0, 0.40}
\definecolor{const}{rgb}{0.40, 0, 0}
\definecolor{type}{rgb}{0, 0.40, 0}
\definecolor{typecheck}{rgb}{0.40, 0, 0.40}
\definecolor{row}{rgb}{0.7,0.95,0.95}
\definecolor{row2}{rgb}{0.95,0.95,0.95}
\definecolor{row3}{rgb}{0.90,0.90,0.90}
\definecolor{mygreen}{rgb}{0,0.5,0}
\definecolor{mygreen2}{rgb}{0,0.25,0}

\section{Background}

\myparagraph{Fuzz Testing}
Fuzz testing, or ``fuzzing'', is a method for automated software testing using a (biased) random search over the input space.
Popular fuzz testing tools, such as \textsc{LibFuzzer}~\cite{libfuzzer} and the \emph{American Fuzzy Lop} (AFL)~\cite{afl}, are configured with a target program $P$ and an initial seed corpus $T$.
During the fuzzing process, new inputs for $P$ are automatically generated using random mutation over the elements of $T$.
Each newly generated input $t$ is then tested against program $P$ to detect \emph{crashes} (e.g., \verb+SIGSEGV+), indicating a bug or security vulnerability.
The fuzzing process can be purely random (e.g., \emph{blackbox} fuzzing) or use information derived from the program $P$ to guide test selection (e.g., \emph{whitebox} or \emph{greybox} fuzzing).
In this paper we focus on greybox fuzzers, such as AFL, which collect \emph{branch coverage information} for each newly generated input $t$.
Inputs which increase branch coverage, i.e., cause the execution of new code branches, are deemed ``interesting'' and will be added into the corpus $T$.
This effectively biases the random search towards inputs that explore more paths, thereby allowing for more bugs to be discovered.

Fuzz testing tools, such as AFL, need to run an instance of the target program $P$ for each newly generated test input $t$.
This is implemented using a \emph{fork server} which is illustrated in Algorithm~\ref{alg:fork}.

\newcommand\mycommfont[1]{\ttfamily\textcolor{blue}{#1}}
\SetCommentSty{mycommfont}
\begin{algorithm}[ht]
\SetAlgoLined
 \While{$\mathit{recvMsg}()$}{
  $\mathit{pid} = \texttt{fork}()$\;
  \eIf{$\mathit{pid} ~==~ 0$}
  {
      \texttt{main();} $~~~$ \tcp{Execute the test case}
  }
  {
      $\texttt{waitpid}(\mathit{pid}, \texttt{\&}\mathit{status})$\;
      $\mathit{sendMsg}(\mathit{status})$\;
  }
 }
 \caption{Fork server loop.\label{alg:fork}}
\end{algorithm}

\noindent
The fork server is injected into the target program during program initialization (i.e., before \verb+main()+ is called).
The fork server essentially implements a simple \emph{Remote Procedure Call} (RPC) loop, where the external fuzzer sends a message for each new input $t$ ready for testing.
This induces the target program to fork,
creating a child process copy of the original (parent) process.
The child process (i.e., where $\mathit{pid}\texttt{~==~}0$) calls \verb+main()+ to execute the test case $t$.
The parent process waits for the child to finish executing, 
and communicates the \emph{exit status} (normal execution or \emph{crash}) back to the external fuzz testing tool.
For a typical fuzz testing application, the target program $P$ will be forked hundreds or thousands of times per second---once for each generated input $t$.

Alternatives to the fork server design exist, such as \emph{in-process fuzzing} used by \textsc{LibFuzzer}~\cite{libfuzzer} or AFL's \emph{persistent mode}.
However, this design requires the developer to manually create a \emph{driver} which guides the fuzzing process, as well as reset the program state between tests.
The fork server design avoids the need for a manual driver and can fully automate the fuzzing process.
Most existing literature~\cite{bohme2017coverage, chen2018angora, aflplusplus, fuzzan, li2017steelix, peng2018t, zhangdebloating} assumes fork-mode fuzzing.

\label{sec:sanitizer}
\myparagraph{Memory Error Sanitizers}
Like fuzz testing tools, the aim of \emph{sanitizers}~\cite{song19sok} is to detect bugs in software. Sanitizers typically use a program transformation and/or a runtime environment modification in order to make bugs more visible.
Many popular sanitizers are implemented as compiler extensions (e.g., an \emph{LLVM Compiler Infrastructure}~\cite{llvm} pass) that insert \emph{instrumentation} to enforce safety properties before critical operations.
In the case of memory error sanitizers, the instrumentation aims to detect memory errors (e.g., buffer overflows, (re)use-after-free), and will be inserted before all memory read and write operations.
Since memory errors will not always cause a crash, a memory error sanitizer is necessary for reliable detection.

Since memory errors are a major source of security vulnerabilities in modern software~\cite{microsoft-bluehat-talk}, the detection of memory errors is of paramount importance.
As such, many different memory error sanitizer designs have been proposed, including~\cite{duck16heap, duck17stack, duck18effective, nagarakatte09softbound, younan10pari, nethercote2007valgrind, duck2022redfat, akritidis09baggy, song19sok, fuzzan, asan, msan}, each with different performance and capabilities.
Each design has its pros and cons in terms of detection capability, performance, and implementation maturity.
A summary of some popular memory error sanitizers is shown in Table~\ref{tab:sanitizers}.
Here, each memory error sanitizer can detect at least one of five classes of memory error (object \emph{overwrites}, \emph{overreads}, \emph{underwrites}, \emph{underreads}, and \emph{use-after-free} errors) over (\emph{heap}, \emph{stack}, and \emph{global}) objects.

Each sanitizer can be differentiated based on memory error detection capability, with some sanitizers being  specialized and others being more general. For example, \emph{stack canaries} are specialized to \emph{stack buffer overwrites}, 
LowFat~\cite{duck16heap, duck17stack} and \emph{Lightweight Bounds Checking} (LBC)~\cite{hasabnis2012light} are specialized to overflows/underflows only,
and FreeSentry~\cite{younan2015free} is specialized to use-after-free errors only.
Furthermore, the detection of memory errors may be partial or imprecise even if supported.
For example, GWP-ASAN~\cite{gwp-asan} only applies protection to randomly selected heap objects, and LowFat/REST~\cite{sinha18rest} tolerate ``small'' overflows that do not intersect with other objects.

Sanitizers such as AddressSanitizer~\cite{asan}
aim to be \emph{general}, and are able to detect all classes of memory error with \emph{byte-level precision}, meaning that even ``small'' overflows will be detected.
To do so, AddressSanitizer implements a form of \emph{memory poisoning} as illustrated in Figure~\ref{fig:poison}.
The basic idea is to mark memory as ``poisoned'' if it can only be accessed using a memory error.
This includes:
\begin{enumerate}[leftmargin=*,topsep=1ex,itemsep=0pt,partopsep=1ex,parsep=1ex]
    \item Poisoning a small \verb+REDZONE+ region that is inserted between each valid allocated object.
    The redzone region is used to detect object bounds overflow/underflow errors.
    \item Poisoning \verb+free()+'ed memory (\verb+FREE+) to detect use-after-free errors.
\end{enumerate}
AddressSanitizer uses a runtime support library to (1) insert and poison redzones between allocated objects, and (2) poison \verb+free()+'ed memory.
In order to detect memory errors, AddressSanitizer instruments all memory access operations to check for poisoned memory:

\newcommand*\rot{\rotatebox{90}}
\newcommand*{\p}[1]{\begin{tikzpicture}[scale=0.08]%
    \draw (0,0) circle (1);
    \fill[fill=black] (0,0) -- (90:1) arc (90:90-#1*3.6:1) -- cycle;
    \end{tikzpicture}}
\begin{table}[tbp]
\setlength{\tabcolsep}{2.5pt}
\footnotesize
\centering
\caption{Comparison of popular memory error sanitizers.
\label{tab:sanitizers}}
\vspace{1em}
\begin{tabular}{|l|ccccc|ccc|cc|ccc|}
\cline{2-14}
\multicolumn{1}{c|}{} &
                   \multicolumn{5}{c|}{\emph{Error Detection}} &
                   \multicolumn{5}{c|}{\emph{Memory}} &
                   \multicolumn{3}{c|}{\emph{Impl.}} \\
\cline{2-14}
\multicolumn{1}{c|}{\emph{Sanitizer}}
                 & \rot{\em Overwrite}
                 & \rot{\em Overread}
                 & \rot{\em Underwrite}
                 & \rot{\em Underread}
                 & \rot{\em Use-after-free~}
                 & \rot{\em Heap}
                 & \rot{\em Stack}
                 & \rot{\em Global}
                 & \rot{\em Locality}
                 & \rot{\em Overhead}
                 & \rot{\em Maintained?$^*$}
                 & \rot{\texttt{x86\_64}\emph{?}}
                 & \rot{\em +Fuzzer?$^\dagger$} \\
\cline{2-13}
\hline
Stack Canaries &
    \p{75} & \p{0} & \p{0} & \p{0} & \p{0} &
    - & \cmark & - & \p{100} & \p{100} &
    \cmark & \cmark & \cmark \\
GWP-ASAN~\cite{gwp-asan} &
    \p{25} & \p{25} & \p{0} & \p{0} & \p{25} &
    \cmark & - & - & \p{100} & \p{75} &
    \cmark & \cmark & - \\
\emph{efence}~\cite{bruce1993electri} &
    \p{75} & \p{75} & \p{0} & \p{0} & \p{100} &
    \cmark & - & - & \p{100} & \p{0} &
    \cmark & \cmark & - \\
FreeSentry~\cite{younan2015free} &
    \p{0} & \p{0} & \p{0} & \p{0} & \p{100} &
    \cmark & \cmark & - & \p{0} & n/a &
    - & - & - \\
LowFat~\cite{duck16heap, duck17stack} &
    \p{75} & \p{75} & \p{100} & \p{100} & \p{0} &
    \cmark & \cmark & \cmark & \p{50} & \p{100} &
    - & \cmark & - \\
REST~\cite{sinha18rest} &
    \p{75} & \p{75} & \p{100} & \p{100} & \p{100} &
    \cmark & \cmark & \cmark & \p{100} & \p{75} &
    - & - & - \\
LBC~\cite{hasabnis2012light} &
    \p{100} & \p{100} & \p{100} & \p{100} & \p{0} &
    \cmark & \cmark & \cmark & \p{50} & \p{50} &
    - & - & - \\
MemCheck~\cite{nethercote2007valgrind} &
    \p{100} & \p{100} & \p{100} & \p{100} & \p{100} &
    \cmark & - & - & \p{0} & \p{0} &
    \cmark & \cmark & - \\
RedFat~\cite{duck2022redfat} &
    \p{100} & \p{100} & \p{100} & \p{100} & \p{100} &
    \cmark & - & - & \p{100} & \p{100} &
    \cmark & \cmark & - \\
AddressSanitizer~\cite{asan} &
    \p{100} & \p{100} & \p{100} & \p{100} & \p{100} &
    \cmark & \cmark & \cmark & \p{0} & \p{0} &
    \cmark & \cmark &  \cmark \\
FuZZan~\cite{fuzzan} &
    \p{100} & \p{100} & \p{100} & \p{100} & \p{100} &
    \cmark & \cmark & \cmark & \p{0} & \p{50} &
    \cmark &  \cmark & \cmark \\
\hline
\hline
\isan &
    \p{100} & \p{100} & \p{100} & \p{100} & \p{100} &
    \cmark & \cmark & \cmark & \p{100} & \p{100} &
    \cmark & \cmark & \cmark \\
\hline
\multicolumn{1}{}{} \\
\end{tabular}

\begin{tabular}{llll}
Key: & \emph{Error Detection} &
    \emph{Mem. Locality} &
    \emph{Mem. Overhead} \\
& $\p{0} = \text{no support}$ &
    $\p{0} = \text{disjoint}$ &
    $\p{0} = \text{high},{>}2{\times}$ \\
& $\p{25} = \text{partial,random}$ &
    $\p{50} = \text{mixed}$ &
    $\p{50} = \text{mixed}$ \\
& $\p{75} = \text{byte imprecise}$ &
    $\p{100} = \text{none/unified}$ &
    $\p{75} = \text{mixed,minimal}$ \\ 
& $\p{100} = \text{byte precise}$ &
    &
    $\p{100} = \text{minimal}$ \\
& & &
    $\text{n/a} = \text{data not avialable}$ \\
\\
\multicolumn{4}{l}{$*$ Maintained? $=$ public repository with recent (${<}1$ year) commits.} \\
\multicolumn{4}{l}{$\dagger$ +Fuzzer? $=$ known public fuzzer integration.} \\
\multicolumn{4}{l}{{\em Impl.} $=$ Implementation feature similarity.}
\end{tabular}
\end{table}

\lstset{language=C,
                basicstyle=\ttfamily\fontseries{m}\selectfont,
                deletekeywords={int},
                keywordstyle=\color{keyword}\ttfamily\fontseries{b}\selectfont,
                keywordstyle=[3]\color{typecheck}\ttfamily\fontseries{b}\selectfont,
                keywordstyle=[4]\color{const}\ttfamily\fontseries{b}\selectfont,
                stringstyle=\color{red}\ttfamily,
                commentstyle=\color{comment}\ttfamily\fontseries{b}\selectfont,
                morekeywords=[3]{poisoned, error},
                morekeywords=[2]{void, uint8_t},
                morekeywords=[4]{Token, NONCE},
}
{\small
\begin{lstlisting}
  if (poisoned(p))          // Instrumentation
    error();
  *p = v; /* or */ v = *p;  // Access
\end{lstlisting}
}

\vspace{-0.1em}
\noindent
Here, $\mathtt{poisoned}(p)$ holds iff the corresponding memory at address $p$ is poisoned.
If so, $\mathtt{error}()$ will report the memory error and abort the program.

Memory poisoning is a popular technique implemented by many different sanitizers, such as~\cite{asan, fuzzan, nethercote2007valgrind, hasabnis2012light, sinha18rest}.
The main distinction is \emph{how} memory poisoning is implemented.
For example, AddressSanitizer implements memory poisoning by dividing the program's virtual address space into two parts:
\emph{application memory} and \emph{shadow memory}.
The shadow memory tracks the (un)poisoned state of each byte of application memory.
To do so, each 8-byte word in application memory is mapped to a corresponding shadow byte as follows:
($\mathit{addr}_\mathit{shadow} = \mathit{offset}_\mathit{shadow}+(\mathit{addr}~/~8)$).
The shadow byte tracks which of the corresponding application bytes have been poisoned, allowing for byte-precise memory error detection.
Shadow memory is a form of \emph{disjoint metadata}---i.e., an additional metadata that is (1) maintained by the sanitizer, and (2) disjoint from the application memory/data.
Disjoint metadata adds memory \emph{overheads} (i.e., the extra space for the shadow memory) and affects memory \emph{locality} (i.e., the application and shadow memory are disjoint).

Alternative implementations of memory poisoning are possible.
One example is \emph{Randomized Embedded Tokens} (RETs) as implemented by REST~\cite{sinha18rest} and LBC~\cite{hasabnis2012light}.
Here, poisoned memory is represented by a special \emph{token} that is initialized to some predetermined random nonce value.
Memory is deemed ``poisoned'' if it directly stores the nonce value:

{\small
\begin{lstlisting}
poisoned(p) = (*(p - p % sizeof(Token))==NONCE)
\end{lstlisting}
}

\vspace{-0.1em}
\noindent
This approach represents the memory (un)poisoned state using the same memory itself.
Since there is no disjoint metadata, both the memory overhead and locality are improved.

That said, a memory poisoning implementation based on RETs may suffer from limitations, such as  \emph{false detections} and a coarse-grained memory error detection \emph{granularity}.
Here, a false detection will occur if the \texttt{NONCE} value happens to collide with a legitimate value created by the program during normal execution, meaning that legitimate access may be flagged as a memory error.
To counter this, REST uses a very large token size (a whole 512 bit cache line) combined with a strong pseudo-random source, meaning that collisions are essentially impossible over practical timescales.
That said, large (multi-byte) tokens introduce a new problem: a reduced memory error detection granularity.
Specifically, due to size constraints, REST only stores tokens in addresses that are a multiple of the token size.
This also means that it is necessary to ``pad'' allocated objects to the nearest token size multiple.
For example, given the 512 bit (64 byte) token size of REST, a call to \verb+malloc(27)+ will:
\begin{enumerate}[topsep=1ex,itemsep=0pt,partopsep=1ex,parsep=0ex]
\item pad the allocation size by $64{-}27{=}37$ bytes,
\item allocate the object aligned a 64 byte boundary, and
\item store a token in bytes $64..127$ to implement the redzone.
\end{enumerate}
This means that any overflow into the padding bytes ($27..63$) will not access the token and will therefore not be detected as a memory error.
This is a \emph{missed detection} (i.e., \emph{false negative}) meaning that REST is not byte-precise.

The idea of randomized tokens is also used by (and first pioneered by) \emph{Lightweight Bounds Checking} (LBC)~\cite{hasabnis2012light}.\footnote{LBC also uses a different terminology, with \emph{guard zone value} used in place of \emph{randomized embedded token}.}
Unlike REST, LBC uses a single byte (8 bit) token size, which allows for byte precise memory error detection, but also means that collisions are inevitable.
To avoid false detections, LBC implements a hybrid approach that retains a disjoint metadata for distinguishing collisions from legitimate memory errors.

\definecolor{redzone}{RGB}{200,0,0}
\begin{figure}[t]
    \centering
    \includegraphics[scale=1.45]{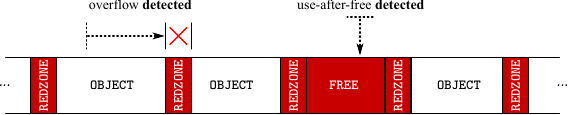}
    \caption{An illustration of memory poisoning.
    Here, (a) each allocated \texttt{OBJECT} is padded with a \emph{poisoned redzone} to detect bounds over/under flows, and
    (b) free'ed memory is also poisoned to detect
    use-after-free errors.
    The memory (un)poisoning operations are implemented using program transformation (for global/stack objects) and a modified memory allocator (for heap objects).
    \label{fig:poison}}
\end{figure}

\myparagraph{Problem Statement: Fuzzing+Sanitizers}
It is natural to combine memory error sanitizers with fuzz testing.
The fuzzer will explore the input space, and the sanitizer will detect ``silent'' memory errors that would otherwise go undetected.
In principle, the fuzz testing and sanitizers should work together synergistically, allowing for the detection of more memory error bugs than either tool alone.
In practice, however, the combination of fuzz testing and sanitizers suffers from poor performance~\cite{fuzzan}.

The root cause of the problem lies with the interaction between the
\emph{copy-on-write} (COW) semantics of the \verb+fork()+ system call, and the
initialization/use of any disjoint metadata by the sanitizer.
The basic idea of COW is to delay the copying of memory by initially sharing all physical pages between the parent and child process.
To do so, all writable memory will be marked as \emph{copy-on-write} by the \verb+fork()+ system call, meaning that a \emph{page fault} will be generated when the corresponding page is first written to by the child process.
This allows the kernel to intercept writes and copy pages \emph{lazily}, which is useful for optimizing the common \verb+fork++\verb+execv+ use-case.
However, in the case of fuzzing+sanitizers, this design can lead to a proliferation of page faults since memory must be modified in two distinct locations:
once for allocated objects and once again for the disjoint metadata.
Since page fault handling is a relatively expensive operation, especially for a disjoint (non-local) address space, this can become the main source of overhead.
In addition to page faults, AddressSanitizer also introduces other sources of overheads relating to \verb+fork()+, such as the copying of kernel data structures including the \emph{Virtual Memory Areas} (VMAs), page tables, and associated teardown costs.
We refer the reader to~\cite{fuzzan} for a more detailed analysis.

One idea would be to choose an alternative memory error sanitizer with lower memory overheads and higher locality.
However, as shown in Table~\ref{tab:sanitizers}, no existing sanitizer design satisfies the dual requirements of optimal memory usage and high memory error detection coverage.
For example, some sanitizers, such as stack canaries, GWP-ASAN, \emph{efence}, LowFat, LBC and REST, achieve excellent/good locality and overheads.
However, this comes at the cost of a reduced memory error detection coverage and/or the lack of support for standard \verb+x86_64+ hardware.
Another idea would be to optimize the disjoint metadata representation.
For example, \fuzzan retains AddressSanitizer's error detection coverage, but can also switch metadata representations to a more compact representation using feedback based on fuzzing performance~\cite{fuzzan}.
Nevertheless, \fuzzan's approach still uses a disjoint metadata, meaning that the overheads are mitigated rather than eliminated altogether.

\subsection{Our Design}\label{sec:design}

Our aim is to optimize fuzzing+sanitizer performance without reducing memory error detection coverage.
To do so, we propose a new sanitizer design based on a variant of \emph{Randomized Embedded Tokens} (RETs) that does not use \emph{shadow map} or other \emph{disjoint metadata} representation.
The key idea is that, by tracking the (un)poisoned state using the memory itself, we avoid any additional page fault (and other overhead) that would be generated by the disjoint metadata initialization and access.
Furthermore, since the instrumented check and corresponding memory operation both access the same memory, the overall number of page faults generated by the instrumented program remains roughly equivalent (i.e., within the order of magnitude) as the uninstrumented program, except for a modest increase due the insertion of redzones and quarantines.
We argue that a RET-based sanitizer design is optimized for memory \emph{locality}, and this improves the performance of fork-based fuzzing.
We summarize the main elements of our design as follows:

\myparagraph{Token Size}
As shown in Table~\ref{tab:sanitizers}, some existing sanitizers
already use a RET-based design.
However, the existing tools suffer from various limitations and are not optimized for fuzzing.
For example, REST~\cite{sinha18rest} uses a very large token size (512 bits) resulting in 
imprecise memory error detection.
In contrast, LBC~\cite{hasabnis2012light} uses a very small token size (8 bits), but must retain a disjoint metadata to avoid false detections.
We argue that, for the application of fuzzing, some small level of false detections can be tolerated provided that the real errors (i.e., ``true'' positives) are not overwhelmed.
We therefore propose a ``medium'' token size of 64 bits which is sufficient to avoid false detections in practice without the need for a separate disjoint metadata.
For example, assuming a 64 bit token size and one billion randomized writes per second, we would expect the first false detection to occur after an average of ${\sim}584.9$ years.

Finally, we note that, for the application of fuzzing, false detections can be also mitigated by re-executing the test case with a different randomized \verb+NONCE+ value.
This process can be automated, similar to how AFL handles false positives due to {\em hangs}/{\em timeouts}.

\myparagraph{Memory Error Detection Granularity}
Another design challenge is the memory error detection granularity.
Under the basic RET-design, tokens are stored on token-size aligned boundaries, which means an 8-byte alignment for 64 bit tokens.
Although this is an improvement over REST, it nevertheless means that
object bounds errors can only be detected within a granularity of 8 bytes.
Small overflow of $1..7$ bytes into object padding will not be detected (unlike tools such as AddressSanitizer which are byte-precise).
To address the issue of granularity, we propose a refinement to the basic RET-based design which additionally encodes object boundary information directly into the token representation itself.
This information can then be retrieved at runtime, and compared against the bounds of the memory access, allowing for fine-grained (byte-precise) detection. 
This enables a similar memory error detection capability compared to the current state-of-the-art memory error sanitizers while still avoiding disjoint metadata.

That said, by encoding boundary information into the token, we must reduce the effective nonce size by 3 bits.
This lowers the expected time to the first false detection from centuries to decades (e.g., to ${\sim}73.1$ years for the example above).
We believe this is still a tolerable false detection rate.

\myparagraph{Hardware}
The final design challenge is a practical implementation.
REST~\cite{sinha18rest} is implemented as a non-standard hardware extension, and LBC~\cite{hasabnis2012light} is specialized to 32bit \verb+x86+ systems only.
In contrast, our RET-based design is the first to target standard hardware (\verb+x86_64+) and standard fuzzers.

\section{Basic Memory Error Checking}\label{sec:basic}

\begin{figure}[t]
\vspace{2mm}
{\footnotesize
\newcommand{\highlightex}[1]{%
    \ifboolexpr{
        test {\ifnumcomp{#1}{=}{2}} or
        test {\ifnumcomp{#1}{=}{3}} or
        test {\ifnumcomp{#1}{=}{4}} or
        test {\ifnumcomp{#1}{=}{5}} or
        test {\ifnumcomp{#1}{=}{6}}
    }{\color{black!15}}{\ifboolexpr{
        test {\ifnumcomp{#1}{=}{9}} or
        test {\ifnumcomp{#1}{=}{10}} or
        test {\ifnumcomp{#1}{=}{11}} or
        test {\ifnumcomp{#1}{=}{12}} or
        test {\ifnumcomp{#1}{=}{13}}
    }{\color{blue!15}}{\color{row2}}}
}
\lstset{language=C,
                frame=single,
                numbers=left,
                numberstyle=\tiny,
                xleftmargin=2em,
                xrightmargin=4pt,
                numbersep=6pt,
                escapechar=@,
                basicstyle=\ttfamily\fontseries{m}\selectfont,
                deletekeywords={int},
                keywordstyle=\color{keyword}\ttfamily\fontseries{b}\selectfont,
                keywordstyle=[2]\color{type}\ttfamily\fontseries{b}\selectfont,
                keywordstyle=[3]\color{typecheck}\ttfamily\fontseries{b}\selectfont,
                keywordstyle=[4]\color{const}\ttfamily\fontseries{b}\selectfont,
                stringstyle=\color{red}\ttfamily,
                commentstyle=\color{comment}\ttfamily\fontseries{b}\selectfont,
                morekeywords=[2]{void, uint64_t,Token},
                morekeywords=[4]{NONCE},
                linebackgroundcolor=\highlightex{\value{lstnumber}},
                mathescape
}
\begin{lstlisting}[escapechar=@]
/* Randomized Embedded Token check */
void *ub = ptr + sizeof(*ptr) - 1;
void *tptr = ub - (ub % sizeof(Token));
Token token = *(Token *)tptr;
if (token.random == NONCE)
    error();

/* Boundary check */
tptr += sizeof(Token);
token = *(Token *)tptr;
if (token.random == NONCE &&
    ub % sizeof(Token) > token.boundary)
    error();

/* The original memory access */
*ptr = val;   @$\mathit{or}$@   val = *ptr;
\end{lstlisting}
}
\caption{Pseudo-code for the
\emph{Randomized Embedded Token} (RET) check and the optional \emph{Refined Boundary} check.
The RET-check is
{\setlength{\fboxsep}{0pt}\colorbox{black!15}{highlighted}}
in lines 2-6, and the boundary-check is
{\setlength{\fboxsep}{0pt}\colorbox{blue!15}{highlighted}}
in lines 9-13.\label{fig:check}}
\end{figure}

\noindent
For ease of presentation, we define \emph{Randomized Embedded Tokens} (RETs) using the following structure type:
\lstset{language=C,
                basicstyle=\ttfamily\fontseries{m}\selectfont,
                deletekeywords={int},
                keywordstyle=\color{keyword}\ttfamily\fontseries{b}\selectfont,
                keywordstyle=[2]\color{type}\ttfamily\fontseries{b}\selectfont,
                keywordstyle=[3]\color{typecheck}\ttfamily\fontseries{b}\selectfont,
                keywordstyle=[4]\color{const}\ttfamily\fontseries{b}\selectfont,
                stringstyle=\color{red}\ttfamily,
                commentstyle=\color{comment}\ttfamily\fontseries{b}\selectfont,
                morekeywords=[2]{void, uint64_t,Token},
}

{\small
\begin{lstlisting}
     struct Token { uint64_t random; };
\end{lstlisting}
}
\noindent
For a value $t$ of type \verb+Token+ to be a valid RET, the \verb+random+ bits must match a predetermined 64 bit randomized constant denoted by the name \verb+NONCE+, i.e.,
$\texttt{(}t\texttt{.random}~\texttt{==}~\texttt{NONCE)}$.
The \verb+NONCE+ constant is initialized once during program initialization using a suitable pseudorandom source.
We define RETs as structures to allow for extensions under the refined design (see Section~\ref{sec:refined}).

\myparagraph{Instrumentation Schema}
As with other memory error sanitizer designs, our underlying approach is to transform the program (e.g., using an \emph{LLVM compiler infrastructure} pass~\cite{llvm}) to insert \emph{instrumentation} before each memory access.
The instrumentation is additional code that checks whether the given \emph{safety property} has been violated or not, and if it has, the program will abort.
In the case of our sanitizer design, the safety property is that the corresponding accessed memory is not \emph{poisoned}.
Later, we combine the instrumentation with a suitable runtime environment that poisons \emph{free} and \emph{redzone} memory in order to enforce memory safety.

The baseline instrumentation schema is
{\setlength{\fboxsep}{0pt}\colorbox{black!15}{highlighted}}
in Figure~\ref{fig:check} lines 2--6, and is inserted before each memory access operation represented by line 16.
Lines 9--13 contain additional instrumentation that we shall ignore for now.
We can define the \emph{range} of a memory access in terms of the \emph{lower bound} ($\mathit{lb}$) and \emph{upper bound} ($\mathit{ub}$):
$$\mathit{lb}..\mathit{ub} = \texttt{ptr}~..~ \texttt{ptr}+\mathtt{sizeof}(\texttt{*ptr})-1$$
The lower and upper bounds are the addresses of the first and last byte accessed by the memory operation.

Figure~\ref{fig:check} line 2 calculates the upper bound.
Line 3 calculates the corresponding \emph{token pointer} (\verb+tptr+) by \emph{aligning} the $\mathit{ub}$ to the nearest token-sized multiple.
This effectively discards the original alignment of $\mathit{ub}$, meaning that any arbitrary overlap between the memory operation and the token can be detected.
Finally, lines 4--6 read the corresponding token from memory and compare the value with the predetermined \verb+NONCE+ constant.
If the values match, the corresponding memory is deemed \emph{poisoned}, and the execution of the program will be \emph{aborted} (line 6).

The instrumentation in Figure~\ref{fig:check} lines 2--6 consists of pointer arithmetic (lines 2--3), memory dereference (line 4), and an error check (lines 5--6).
Importantly, the memory dereference (line 4) will only access memory that is to be accessed anyway by the memory operation (line 16).
In other words, the line 4 memory dereference will not generate an extra page fault that would not have been generated by line 16 anyway.
In addition to line 4, the error check (lines 5--6) also accesses memory to retrieve the \verb+NONCE+ value that is stored in a global variable.
Since the \verb+NONCE+ is stored in a single location, this will generate at most one additional page fault under normal conditions, so can be treated as a once-off cost.

\myparagraph{Runtime Support}
To enforce \emph{memory safety}, the runtime environment is also modified in order to \emph{poison} both \emph{redzone} and \emph{free} memory, thereby allowing the instrumentation to detect the error.
Each class of object (heap/stack/global) is handled differently.

\paragraph{Heap Allocated Objects}
Standard heap allocation functions, i.e., \verb+malloc+, \verb+realloc+, \verb+new+, etc., are replaced by new versions that place \emph{redzone}s around each allocated object.
The process is similar to how redzones are implemented in other memory error sanitizers, such as AddressSanitizer, except:
\begin{itemize}[leftmargin=*,topsep=1ex,itemsep=0pt,partopsep=1ex,parsep=0ex]
\item[-] Poisoning is implemented by writing a \verb+NONCE+-initialized token directly into redzone memory.
\item[-] The redzone size is 1 or 2 tokens (depending on alignment).
\item[-] The redzone is placed at the end of the object.
Underflows are detected using the redzone of the previous object allocated in memory.
\end{itemize}
For heap objects, we have implemented a simple custom memory allocator that allocates objects \emph{contiguously}, i.e., there are no gaps between objects except for redzones.

For heap \emph{deallocation}, the free'ed object is poisoned by filling the corresponding memory with a \verb+NONCE+-initialized token.
Any subsequent access to the object will therefore be detected as an error, i.e., \emph{use-after-free} error.
To help mitigate \emph{reuse-after-free} errors (i.e., if a dangling pointer is accessed after the underlying memory was reallocated)
our solution also implements a \emph{quarantine} for free'ed objects.
A quarantine is essentially a queue for free'ed objects which aims to delay reallocation, making it more likely that reuse-after-free errors will be detected.
After an object is removed from the quarantine in order to be reallocated, the corresponding memory will be \emph{zeroed} to ``\emph{unpoison}'' the memory before use.

\paragraph{Stack Allocated Objects}
Stack allocated objects are handled using a program transformation implemented as an LLVM~\cite{llvm} pass (similar to the instrumentation pass).
To add redzones to stack objects, the allocation size is first modified to include space for both the original object as well as the redzone memory.
The augmented object is then allocated from the stack as per usual, and the redzone memory is poisoned by writing a \verb+NONCE+-initialized token, as with the case with heap-allocated objects.
The remaining memory is zeroed to remove any residual token values that may be leftover from previous stack allocations.

\paragraph{Global Objects}
Global objects are similarly implemented using an LLVM~\cite{llvm} pass.
The idea is the same:
the object size is extended to include some additional space for redzone memory, which is poisoned using a \verb+NONCE+-initialized token.

\section{Refined Boundary Checking}\label{sec:refined}

The basic RET-check of Figure~\ref{fig:check} can protect object bounds overflow errors up to a granularity of the \emph{token size}, i.e., $\texttt{sizeof}(\texttt{Token}){=}8$ bytes.
Since embedded tokens must be aligned to the token size, allocated objects must therefore be \emph{padded} to the nearest 8-byte boundary (see Section~\ref{sec:design}).
To address the issue of granularity, we propose a refinement of the original \emph{Randomized Embedded Token} (RET) design.
The basic idea is to encode \emph{object boundary information} into embedded tokens in addition to the randomized \verb+NONCE+.
This boundary information can be retrieved at runtime and checked against the bounds of the memory access.
The refined token design therefore consists of two components:
\begin{itemize}[leftmargin=*,topsep=1ex,itemsep=0pt,partopsep=1ex,parsep=0ex]
\item[-] \emph{random}: The \verb+NONCE+ value, same as before.
\item[-] \emph{boundary}: An encoding of the \emph{object boundary} in the form:
$$\mathit{size} \bmod \mathtt{sizeof}(\texttt{Token})$$ where $\mathit{size}$ is the object size.
\end{itemize}
Conceptually, the refined token design is represented by a structure with two bitfields:

\lstset{language=C,
                basicstyle=\ttfamily\fontseries{m}\selectfont,
                deletekeywords={int},
                keywordstyle=\color{keyword}\ttfamily\fontseries{b}\selectfont,
                keywordstyle=[2]\color{type}\ttfamily\fontseries{b}\selectfont,
                keywordstyle=[3]\color{typecheck}\ttfamily\fontseries{b}\selectfont,
                keywordstyle=[4]\color{const}\ttfamily\fontseries{b}\selectfont,
                stringstyle=\color{red}\ttfamily,
                commentstyle=\color{comment}\ttfamily\fontseries{b}\selectfont,
                morekeywords=[2]{void, uint64_t,Token},
}
{\small
\begin{lstlisting}
 struct Token {
    uint64_t random:61;  // NONCE
    uint64_t boundary:3; // Boundary encoding
 };
\end{lstlisting}
}

\noindent The \emph{boundary} field must be at least $\log_2 \mathtt{sizeof}(\texttt{Token}) = 3$ bits in order to represent all possible boundary values.
The \emph{random} field has also been reduced to 61 bits (from 64).

\begin{figure}[t]
    \centering
    \includegraphics[scale=1.1]{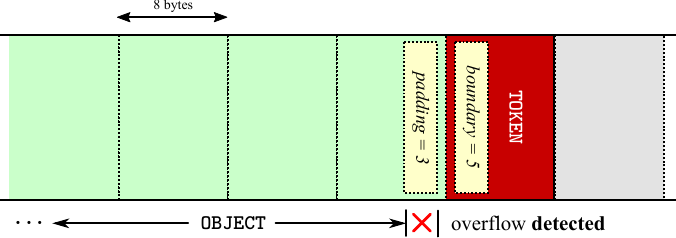}
    \caption{Example of accurate overflow detection with \emph{boundary checking}.
    Here, in addition to the random value, the token also encodes the object \emph{boundary} modulo the token size,
    i.e., $\mathit{boundary}{=}\mathtt{sizeof}(\texttt{Token}){-}\mathit{padding}$
    Any access into \emph{padding} can therefore be detected by comparing the offset (modulo the token size) with the encoded boundary.
    \label{fig:boundary}}
\end{figure}

In order to detect overflows into padding, memory access must be instrumented with an additional \emph{boundary check}.
The basic idea is illustrated in Figure~\ref{fig:boundary}.
As before, we assume the allocated \texttt{OBJECT} is immediately followed by a redzone poisoned by a \verb+NONCE+-initialized token.
In this example, we assume the object $\mathit{size}$ is not a perfect multiple of the token size, e.g., $(\mathit{size} \bmod \mathtt{sizeof}(\texttt{Token})) = 5$, meaning that an additional $\mathtt{sizeof}(\texttt{Token}){-}5 = 3$ bytes of padding is used.
Overflows into the padding will not be detected by the basic RET-check alone, since the padding is too small to store a token.
To detect such overflows, we instrument the memory access with an additional \emph{boundary check} that:
\begin{enumerate}[leftmargin=*,topsep=1ex,itemsep=0pt,partopsep=1ex,parsep=0ex]
    \item Examines the \emph{next} word in memory from the \emph{current} word that is to be accessed;
    \item If the \emph{next} word is \textbf{not} a token (i.e., the \emph{random} bits do not match the \verb+NONCE+), then the memory access is allowed;
    \item Otherwise if the next word is a token, then we retrieve the \emph{boundary} field and compare it against the memory access range $\mathit{lb}..\mathit{ub}$.
    A memory error detected if the following (\textsc{Boundary-Check}) condition does \textbf{not} hold:
    \begin{align*}
        (\mathit{ub} \bmod \mathtt{sizeof}(\texttt{Token})) \leq \mathit{boundary} 
    \end{align*}
\end{enumerate}
Note that, by examining the \emph{next} word in memory, we essentially bypass the problem of the padding having insufficient space.
In the example from Figure~\ref{fig:boundary}, any memory access that overlaps with the \emph{padding} will not satisfy (\textsc{Boundary-Check}), and will therefore be detected.
All other access within the object will be deemed valid, either because the next word in memory is not a token, or the token \emph{boundary} is consistent with the memory access range.

\myparagraph{Instrumentation Schema}
The boundary-check instrumentation is
{\setlength{\fboxsep}{0pt}\colorbox{blue!15}{highlighted}}
in Figure~\ref{fig:check} lines 9--13.
Here, we assume that the memory access (line 16) has already passed the RET-check (lines 2--6), meaning that the \emph{current} word pointed to by the memory access \emph{upper bound} ($\mathit{ub}$) does \emph{not} contain a token.
However, it is still possible that the current word contains the object/padding boundary, and that the $\mathit{ub}$ exceeds this boundary (an overflow error).
The purpose of the \emph{boundary check} is to detect this case.

In order to complete the boundary check, the \emph{next} word in memory must be examined.
Here, lines 9--10 load the \emph{next} word after the upper bound ($\mathit{ub}$) into the $\texttt{token}$ variable.
Line 11 compares
$\texttt{token.random}$ against the \verb+NONCE+, and if there is a match, the following information can be deduced:
\begin{itemize}[leftmargin=*,topsep=1ex,itemsep=0pt,partopsep=1ex,parsep=0ex]
\item[-] The \emph{next} word is part of the \emph{redzone} of the current object; 
\item[-] The \emph{current} word contains the object/padding boundary which is encoded in the $\texttt{token.boundary}$ field.
\end{itemize}
Line 12 therefore checks whether the upper bound ($\mathit{ub}$) exceeds the $\texttt{token.boundary}$ value, and if it does, an error is reported and execution is aborted (line 13).

The next word may reside in a different page, which may be inaccessible.
This could be handled by disabling the refined check over page boundaries, for a slight loss of precision.
Alternatively, all mappings could be extended by a \verb+NONCE+-initialized page.
This can be implemented lazily, by using a signal handler to detect boundary-check induced faults, and then extending the corresponding mapping ``on-demand''.

\section{Experimental Setup}\label{sec:set}

We experimentally validate our sanitizer design in terms of error detection capability, performance, coverage, bug finding capability, flexibility, as well as false detections.
In this section, we give an overview of the experimental setup.

We have implemented our design in the form of the \emph{REt+fuZZing +sANitzer} (\isan) for the \verb+x86_64+.
We will evaluate two main configurations of \isan:
\begin{itemize}[leftmargin=*,topsep=1ex,itemsep=0pt,partopsep=1ex,parsep=0ex]
    \item[-] \isan: Fine-granularity memory error detection with both RET (Section~\ref{sec:basic}) and byte-accurate boundary checking (Section~\ref{sec:refined}); and 
    \item[-] \bisan: Reduced-granularity memory error detection with RET-checking only.
    This version is faster but may not detect some overflows into object padding.
\end{itemize}
Our \isan implementation comprises two parts: an \emph{LLVM Compiler Infrastructure} (LLVM)~\cite{llvm} pass and runtime library.
The \isan LLVM pass:
\begin{itemize}[leftmargin=*]
    \item[-] Transforms all memory operations (e.g., \verb+load+/\verb+store+) to insert RET and boundary-checking instrumentation.
    In the case of \bisan, the boundary checking is omitted.
    \item[-] Transform all stack allocations (e.g., \verb+alloca+) and global objects to new versions that are protected by redzones.
\end{itemize}
The \isan runtime library
implements replacement heap allocation functions
(e.g., \verb+malloc+, \verb+free+, etc.)
which insert redzones as well as poisons free'ed memory.

\myparagraph{Research Questions}
Our main hypotheses are that a RET-based sanitizer design can (1) exhibit a lower performance overhead under fuzz testing environments, and (2) achieve a similar memory error detection capability as more traditional sanitizer designs, such as AddressSanitizer (\asan).
We investigate these hypotheses, from the performance and effectiveness perspectives, with the following research questions:

\begin{description}
  \item[RQ.1 (Detection Capability)]
  Does \isan detect the same class of memory errors as \asan?
  
  \item[RQ.2 (Execution Speed)] How much faster is \isan compared to \asan under fuzz testing environments?

  \item[RQ.3 (Branch Coverage)] How does the branch coverage for \textsc{ReZZan} compare to \asan?
  
  \item[RQ.4 (Bug Finding Effectiveness)] How much faster can \textsc{ReZZan} expose bugs compared to \asan?

  \item[RQ.5 (Flexibility)] Can \isan be used to fuzz huge programs? Is \isan compatible with other fuzzers? 
  \item[RQ.6 (False Detections)] What is the false detection rate of \textsc{ReZZan} in real execution environments? 
\end{description}

\label{sec:infrastructure}
\myparagraph{Infrastructure}
We run our experiments on an Intel Xeon CPU E5-2660v3 processor with 28 physical and 56 logical cores clocked at 2.4GHz.
Our test machine uses Ubuntu 16.04 (64 bit) LTS with 64GB of RAM,
and a maximum utilization of 26 cores.

For the baseline, we compare against \asan (LLVM-12) and \fuzzan (with dynamic metadata structure switching mode).
Both \asan and \fuzzan are: (1) maintained, (2) support the \verb+x86_64+, and (3) have existing fuzzer integration (see Table~\ref{tab:sanitizers}).
As far as we are aware, \isan is the first RET-based design to support the \verb+x86_64+ as well as being integrated with a fuzzer.
For the fuzzing engine, we use AFL (v2.57b), which is (1) the base of most modern fuzzers, and (2) supported by all sanitizers used in the evaluation.
For the memory error detection capability experiments~(RQ1), we use the Juliet~\cite{juliet} benchmark suite. 
Juliet is a collection of test cases containing common vulnerabilities based on a \emph{Common Weakness Enumeration} (CWE), including heap-buffer-overflow, use-after-free, etc.
For the execution speed and branch coverage experiments~(RQ2, RQ3), we use \verb+cxxfilt+, \verb+nm+, \verb+objdump+, \verb+size+ (all from binutils-2.31), \verb+file+ (from coreutils version 5.35), \verb+jerryscript+ (version 2.4.0), \verb+mupdf+ (version 1.19.0),  \verb+libpng+ (version 1.6.38), \verb+openssl+ (version 1.0.1f), \verb+sqlite3+ (version 3.36.0), and \verb+tcpdump+ (version 4.10.0). 
Our test subjects are widely used by recent fuzzing works~\cite{peng2018t, chen2018angora, bohme2017coverage, li2017steelix, aschermann2019redqueen} as well as \fuzzan~\cite{fuzzan}.
For the bug finding effectiveness experiments~(RQ4), we use Google's fuzzer-test-suite\footnote{\url{https://github.com/google/fuzzer-test-suite}}, which provides a collection of subjects and bugs for testing fuzzing engines.
We use the same bugs as studied by \cite{fuzzan}.
For the initial seed corpus in~(RQ2, RQ3)
we use the same number of valid inputs.
For the bug finding effectiveness experiments~(RQ4), we use the single input provided by the fuzzer-test-suite, or else an empty file if no input is provided.
Each trial is run for 24 hours and repeated 20 times 
(the reported result takes the average).

\section{Evaluation Results}\label{sec:eva}

In this section, we present the experimental results for the six research questions from Section~\ref{sec:infrastructure}.

\myparagraph{RQ.1 Detection Capability}
To evaluate the detection capability of \isan and \bisan against memory errors, we select the following tests from the Juliet test suite: stack buffer overflow (CWE: 121), heap buffer overflow (CWE: 122), buffer underwrite (CWE: 124), buffer overread (CWE: 126), buffer underread (CWE: 127), and use-after-free (CWE: 416).
We exclude test cases where the bug is triggered by data from a socket, standard input, or are only triggered under 32 bit operating system environments.
Some other test cases use a random value that determines whether the bug should be triggered.
For these cases, we fix the value to ensure that the bug is always triggered. 
Each test case provides a \emph{bad} and a \emph{good} function.
The \emph{bad} function will trigger the bug whereas the \emph{good} function will not, allowing for the detection of false negatives and positives respectively.
Since \fuzzan has the same detection capability as \asan~\cite{fuzzan}, we focus our evaluation on \asan only.

\begin{table}[tbp]
\centering\footnotesize
\caption{Detection capability based on the bad test cases (bug triggered).
This checks for false negatives.} 
\label{tab:julietbad}
\begin{tabular}{r|r|r|r|r}
\textbf{CWE (ID)[Bad]} & \textbf{Total}  & \textbf{\asan}  & \textbf{\isan}  & \textbf{\bisan} \\ \hline 
Stack Buffer Overflow (121)   & 2860 & 2856 & 2860 & 2380     \\
Heap Buffer Overflow (122)    & 3246 & 3189 & 3246 & 2724   \\ 
Buffer Underwrite (124)       & 928 & 928 & 890 & 890 \\
Buffer Overread (126)         & 630 & 610 & 630 & 630 \\
Buffer underread (127)        & 928 & 928 & 880 & 880 \\
Use After Free (416) & 392 & 392 & 392 & 392 \\ 
\hline
\multicolumn{1}{r}{\textbf{Pass rate:}} & \multicolumn{1}{r}{} & \multicolumn{1}{r}{99.10\%} & \multicolumn{1}{r}{99.04\%} & \multicolumn{1}{r}{87.89\%} \\

\end{tabular}
    
\end{table}

\begin{table}[tbp]
\centering\footnotesize
\caption{Detection capability based on the good test cases (bug is not triggered).
This checks for false positives.} 
\label{tab:julietgood}
\begin{tabular}{r|r|r|r|r}
\textbf{CWE (ID)[Good]} & \textbf{Total}  & \textbf{\asan}  & \textbf{\isan}  & \textbf{\bisan} \\ \hline 
Stack Buffer Overflow (121)   & 2860 & 2860 & 2860 & 2860  \\ 
Heap Buffer Overflow (122)   & 3246 & 3246 & 3246 & 3246   \\ 
Buffer Underwrite (124)       & 928 & 928 & 928 & 928 \\
Buffer Overread (126)         & 630 & 630 & 630 & 630 \\
Buffer underread (127)        & 928 & 928 & 928 & 928 \\
Use After Free (416) & 392 & 392 & 392 & 392 \\ 
\hline
\multicolumn{1}{r}{\textbf{Pass rate:}} & \multicolumn{1}{r}{} & \multicolumn{1}{r}{100.00\%} & \multicolumn{1}{r}{100.00\%} & \multicolumn{1}{r}{100.00\%} \\
\end{tabular}
    
\end{table}

\autoref{tab:julietbad} shows the evaluation results of \asan, \bisan, and \isan on all \emph{bad} test cases.
The results show that \isan and \asan have a similar error detection capability, at 99.04\% and 99.10\% respectively.
There is a slight deviation due to implementation differences, such as library support and redzone size.
In contrast, the memory error detection capability of \bisan is somewhat reduced, at 87.89\%.
This reduction is expected since \bisan does not support byte-accurate overflow detection into the padding, meaning that some test cases with (e.g., with off-by-one overflows) will not be detected.
As will be seen, \bisan trades error detection capability for greater fuzzing throughput.
\isan and \bisan also perform slightly worse than \asan for underflow detection (CWE 124 and 127).
However, this is because \asan uses a double-wide (32byte) redzone for stack objects by default.
When \isan is similarly configured, both \isan and \asan can detect 100\% of all underflow errors.

\autoref{tab:julietgood} shows the results of \asan, \bisan, and \isan on all the \emph{good} test cases.
For these results, we see that all of \asan, \bisan, \isan pass all test cases. 

\result{For memory error bugs (CWE 121, 122, 124, 126, 127, 416) in the Juliet test suite, \isan passes 99.04\% and \bisan passes 87.89\% of the \emph{bad} test cases respectively.}

\begin{figure}
    \includegraphics[width=\columnwidth,clip]{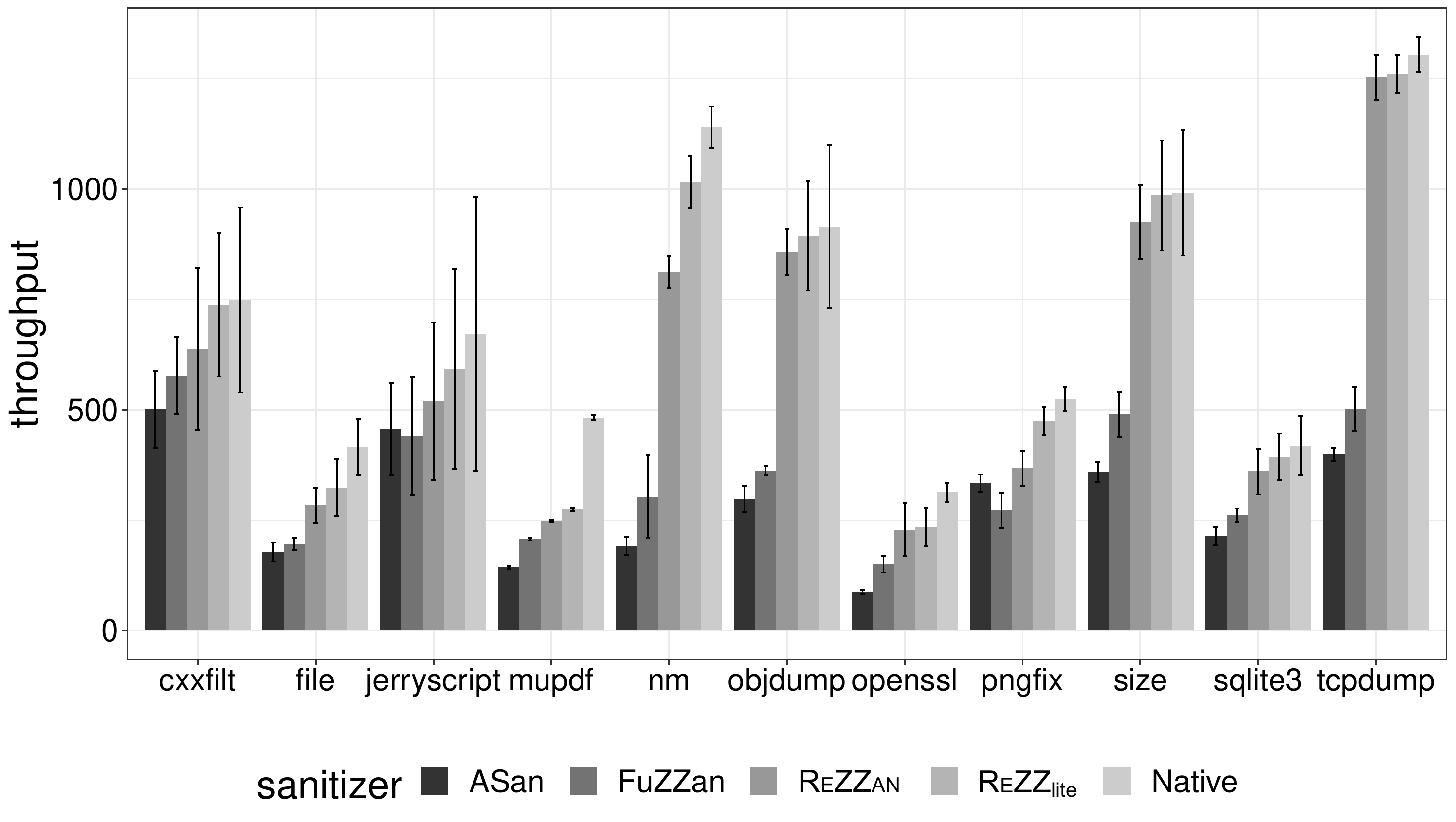}
    \caption{The average throughput (execs/sec) of AFL with four sanitizers and one without (Native).}
    \label{fig:speed}
\end{figure}

\begin{table*}[htbp]
\caption{The average throughput (execs/sec) of AFL with four sanitizers and one without (Native).  Each value is averaged over 20 trials and 24 hours. The percentages in the brackets represent the performance loss for each sanitizer compared to the Native campaign. $\hat A_{12}$ represents the Vargha Delaney A measure, $U$ represents Wilcoxon signed rank, and \emph{Improv.} (Improvement) is the throughput gain of \isan versus \asan.} 
\label{tab:speed}
\centering\small
\begin{tabular}{r|rrrrr|rrr}

    \multirow{2}{*}{\textbf{Subject}} & \multirow{2}{*}{\textbf{\asan}} & \multirow{2}{*}{\textbf{\fuzzan}} & \multirow{2}{*}{\textbf{\isan}} & \multirow{2}{*}{\textbf{\bisan}} & \multirow{2}{*}{\textbf{Native}} & \multicolumn{3}{c}{\textbf{\isan vs. \asan}}\\
    & & & & & & \textbf{$\hat A_{12}$} & \textbf{$U$} & \textbf{Improv.}\\\hline

cxxfilt	& 500.56 (-33.12\%)	& 577.38 (-22.85\%) &	637.25 (-14.85\%) 	& 737.38 (-1.47\%)	& 748.39 (0.00\%) & 0.76 & ${<}$0.01 & 27.31\% \\ 
file	& 177.46 (-57.28\%)	& 195.87 (-52.85\%) &	282.80 (-31.92\%) 	& 323.17 (-22.20\%)	& 415.41 (0.00\%) & 0.95 & ${<}$0.01 & 59.36\% \\ 
jerryscript & 456.74 (-31.98\%) & 440.45 (-34.41\%) & 519.13 (-22.69\%) & 592.16 (-11.82\%) & 671.50 (0.00\%) & 0.61 & 0.24 & 13.66\%\\
mupdf & 142.96 (-70.36\%) & 205.93 (-57.31\%) & 247.60 (-48.65\%) & 273.66 (-43.27\%) & 482.39 (0.00\%) & 1.00 & ${<}$ 0.01 & 73.26\%\\
nm	    & 190.46 (-83.29\%)	& 303.46 (-73.38\%) &	811.71 (-28.79\%) 	& 1016.20 (-10.86\%)	& 1139.96 (0.00\%) & 1.00 & ${<}$0.01 & 326.18\% \\ 
objdump	& 297.51 (-67.48\%)	& 360.95 (-60.54\%) &	857.65 (-6.24\%)     & 893.54 (-2.32\%)	& 914.76 (0.00\%) & 1.00 & ${<}$0.01 & 188.28\% \\ 
openssl & 87.27 (-72.09\%) & 150.00 (-52.04\%) & 228.83 (-26.83\%) & 233.39 (-25.37\%) & 312.73 (0.00\%) & 1.00 & ${<}$0.01 & 162.20\% \\ 
libpng	& 332.89 (-36.53\%)	& 272.65 (-48.01\%) &	366.46 (-30.13\%) 	& 473.54 (-9.71\%)	& 524.46 (0.00\%) & 0.86 & ${<}$0.01 & 10.08\% \\ 
size	& 358.46 (-63.85\%)	& 490.08 (-50.57\%) &	924.95 (-6.71\%)     & 985.67 (-0.59\%)	& 991.49 (0.00\%) & 1.00 & ${<}$0.01 & 158.03\% \\ 
sqlite3 & 213.43 (-49.03\%) & 260.59 (-37.77\%) & 359.67 (-14.11\%) & 393.12 (-6.12\%) & 418.77 (0.00\%) & 1.00 & ${<}$0.01 & 68.52\%\\
tcpdump	& 398.67 (-69.41\%)	& 501.49 (-61.52\%) &   1253.28 (-3.85\%)     & 1260.82 (-3.27\%)	& 1303.40 (0.00\%) & 1.00 & ${<}$0.01 & 214.37\% \\ 
\hline
\multicolumn{1}{l}{\textbf{Avg Loss:}} & -57.67\% & -50.11\% & -21.34\% & -12.45\% & & \multicolumn{2}{r}{\textbf{Avg:}} & 118.30\% \\
\end{tabular}
\end{table*}

\myparagraph{RQ.2 Execution Speed}
The design of \isan has been specifically optimized for performance under fuzz testing environments.
To measure the performance, we conduct a fuzzing campaign with AFL+\isan against the 11 real-world subjects listed in Section~\ref{sec:infrastructure}.
We also compare the performance against \asan~\cite{asan}, \fuzzan~\cite{fuzzan}, as well as the ``Native'' AFL performance with no sanitizer.

The results are shown in \autoref{tab:speed} and illustrated in \autoref{fig:speed}.
Here, we arrange the results from lowest to highest throughput.
Overall, we see that \asan has the lowest executions per second, with a 57.67\% reduction of throughput ($2.36{\times}$ overhead) compared to Native.
Under our experiments, \fuzzan further improves the \asan throughput, with a 50.11\% reduction of throughput ($2.00{\times}$ overhead) compared to Native.
In contrast, both \isan and \bisan show a significantly improved throughput reduction.
Under the coarse-grained \bisan configuration, which can still detect most memory errors including overflows into adjacent objects, the observed throughput reduction is 12.45\% over Native ($1.14{\times}$ overhead).
Even with the byte-accurate \isan configuration, which uses a more complicated instrumentation, the throughput reduction is a relatively modest 21.34\% over Native ($1.27{\times}$ overhead).
To compare against \asan, \autoref{tab:speed} also includes a statistical comparison.
Here, $\hat A_{12}$ is the Vargha Delaney value measuring \emph{effect size}~\cite{vargha2000critique} and $U$ is the Wilcoxon rank sum test.
With $U{<}0.05$, we see that \isan outperforms \asan with statistical significance.

\begin{table}[htbp]
\setlength{\tabcolsep}{4pt}
\centering\footnotesize
\caption{Average page faults over 1000 runs. The \emph{Factor} compares \isan to \asan. } 
\label{tab:pagefault}
\begin{tabular}{r|rrrrr|r}
 \textbf{Subject} & \textbf{\asan} & \textbf{\fuzzan} & \textbf{\isan}  & \textbf{\bisan} & \textbf{Native} &  \textbf{Factor} \\ \hline 
cxxfilt	    & 2893.54	& 2962.79	& 195.57	& 195.49	& 98.55	    & 14.80  \\
file	    & 4131.51	& 3621.84	& 437.95	& 564.88	& 253.24	& 9.43  \\
jerryscript & 3152.13   & 3019.65   & 176.15    & 176.75    & 109.67    & 17.89 \\
mupdf       & 4475.74   & 4423.24   & 647.00    & 653.71    & 305.01    & 6.92  \\
nm	        & 3328.70	& 3293.03	& 273.66	& 277.02	& 128.81	& 12.16  \\
objdump	    & 3416.45	& 3474.61	& 294.21	& 294.19	& 133.71	& 11.61  \\
openssl     & 4692.02   & 1349.85   & 362.16    & 373.14    & 227.90    & 12.96  \\
libpng	    & 4089.07	& 3691.06	& 1204.38	& 1207.24	& 914.74	& 3.40  \\
size	    & 3348.35	& 3251.60	& 491.02	& 494.61	& 129.01	& 6.82  \\
sqlite3     & 3515.80   & 3464.81   & 282.90    & 283.89    & 71.00     & 12.43 \\
tcpdump	    & 4007.29	& 4023.99	& 420.97	& 425.87	& 232.09	& 9.52  \\

\hline
\multicolumn{5}{c}{} & \multicolumn{1}{r}{\textbf{Avg:}} & 10.72${\times}$\\
\end{tabular}
\end{table}

\myparagraph{Page faults}
\isan is specifically designed to optimize the startup/teardown costs caused by the sanitizer~\cite{fuzzan}.
One major source of overhead are page faults arising from the interaction between the \emph{copy-on-write} (COW) semantics of \verb+fork()+ and disjoint metadata.
To quantify the impact, we randomly choose 1000 inputs of each subject generated from the experiments in \autoref{tab:speed} and measure the average number of page faults for each sanitizer.
The results are shown in \autoref{tab:speed}.
Overall we see that the number of page faults is greatly reduced, with a $10.72{\times}$ reduction over \asan on average, and is comparable to the Native execution.
These results are reflected in \autoref{tab:speed}, and validate the RET-based design of \isan in the context of fuzz testing.

\result{When combined with fuzz testing, the overheads of \isan ($1.27{\times}$) and \bisan ($1.14{\times}$) are lower than that of traditional sanitizers \asan ($2.36{\times}$) and \fuzzan ($2.00{\times}$).
The performance of \isan and \bisan is comparable to fuzz testing without any memory error sanitization, as are the number of page faults.}

\myparagraph{RQ.3 Branch Coverage}
Greybox fuzzing aims to increase the branch \emph{coverage} since this can lead to the discovery of new bugs.
For the same setup, 
a higher fuzzer throughput 
ought to translate into higher code coverage as more tests can be explored for the same time budget.
\autoref{tab:path} shows the average branch coverage achieved by the fuzzing campaigns with all 4 sanitizers and without (Native).
We used the \texttt{gcov}\footnote{\url{https://man7.org/linux/man-pages/man1/gcov.1.html}} to measure the branch coverage on the inputs generated in the RQ2.
For all benchmarks, the branch coverage 
for \isan and \bisan are close to that of Native execution, with the coarse-grained \bisan achieving a slightly higher coverage due to greater throughput.
The results for \asan and \fuzzan generally show lower coverage, with \fuzzan performing similarly to \asan.

\result{On average, \isan and \bisan achieve branch coverage similar to Native.
The fuzzing campaign with \isan explores 5.54\% more code branches than \asan within 24 hours.}

\begin{table}[]
\footnotesize
\setlength{\tabcolsep}{1.5pt}
\caption{
    The average branch coverage achieved by each fuzzing campaign with four sanitizers and one without.
    } 
\label{tab:path}
\begin{tabular}{r|rrrrr|rrr}
    \multirow{2}{*}{\textbf{Subject}} & \multirow{2}{*}{\textbf{\asan}} & \multirow{2}{*}{\textbf{\fuzzan}} & \multirow{2}{*}{\textbf{\isan}} & \multirow{2}{*}{\textbf{\bisan}} & \multirow{2}{*}{\textbf{Native}} & \multicolumn{3}{c}{\textbf{\isan vs. \asan}}\\
    & & & & & & \textbf{$\hat A_{12}$} & \textbf{$U$} & \textbf{Impr.}\\\hline
    
cxxfilt	& 1284.95	& 1285.90	& 1287.90	& 1290.35	& 1292.95 & 0.53 & 0.78 & 0.23\% \\
file	& 1393.50	& 1401.42	& 1453.80	& 1516.60	& 1548.00 & 0.52 & 0.85 & 4.33\% \\
jerry	& 8318.90	& 8295.15	& 8434.85	& 8440.80	& 8485.15 & 0.65 & 0.11 & 1.39\% \\
mutool	& 2642.25	& 2629.70	& 2656.70	& 2661.40	& 2676.35 & 0.77 & ${<}$0.01 & 0.55\% \\
nm	    & 1938.75	& 1938.40	& 2206.00	& 2228.70	& 2260.10 & 1.00 & ${<}$0.01 & 13.78\% \\
objdump	& 1292.25	& 1296.53	& 1348.00	& 1300.85	& 1352.70 & 1.00 & ${<}$0.01 & 4.31\% \\
openssl	& 2944.70	& 2415.90	& 3344.50	& 3363.40	& 3373.85 & 0.64 & 0.15 & 13.58\% \\
libpng	& 1939.80	& 1911.60	& 1940.15	& 1936.55	& 1941.10 & 0.68 & 0.06 & 0.02\% \\
size	& 1273.70	& 1293.65	& 1310.50	& 1320.15	& 1328.75 & 0.95 & ${<}$0.01 & 2.89\% \\
sqlite3	& 11261.05	& 11088.80	& 11369.30	& 11401.25	& 11585.28 & 0.58 & 0.41 & 0.96\% \\
tcpdump	& 5960.45	& 5665.60	& 7086.95	& 7168.15	& 7171.70 & 1.00 & ${<}$0.01 & 18.90\% \\

\hline
\multicolumn{7}{c}{} & \multicolumn{1}{r}{\textbf{Avg:}} & 5.54\% \\
\end{tabular}
    
\end{table}

\myparagraph{RQ.4 Bug Finding Effectiveness}
Since \isan/\bisan have lower performance overheads and higher coverage, they ought to be more effective in finding bugs.
To test our hypothesis, we use the same benchmark from~\cite{fuzzan}, which consists of 6 errors chosen from the Google fuzzer-test-suite~\cite{fuzz-test}.
Here, each test is a C program/library containing a bug, including:
\verb+c-ares+ (CVE-2016-5180), 
\verb+libxml2+ (CVE-2015-8317), \verb+openssl+ (A: heartbleed, and B: CVE-2017-3735), and \verb+pcre2+.
The \verb+json+ bug triggers an assertion failure, so is not a memory error.
Nevertheless, it is interesting to include non-memory error bugs which may also benefit from greater throughput and branch coverage.

\begin{table}[htbp]
\setlength{\tabcolsep}{3.2pt}
\centering\footnotesize

\caption{The average time (in seconds) needed to expose the corresponding bug in the Google fuzzer test suite, averaged over 20 trials.
    The libxml2 and openssl (B) benchmarks sometimes exceed the 24 hour timeout, leading to a partial result (*) averaged over 8 and 4 successful trials respectively.
    Here, ($-$) means the bug was not exposed in any run.}
\label{tab:bug}
\begin{tabular}{@{}r|rrrr|rrr@{}}
    \multirow{2}{*}{\textbf{Subject}} & \multirow{2}{*}{\textbf{\asan}} & \multirow{2}{*}{\textbf{\fuzzan}} & \multirow{2}{*}{\textbf{\isan}} & \multirow{2}{*}{\textbf{\bisan}} & \multicolumn{3}{c}{\textbf{\isan vs. \asan}}\\
    & & & & & \textbf{$\hat A_{12}$} & \textbf{$U$} & \textbf{Factor}\\\hline
    c-ares & 80.00 & 47.65 & 22.65 & 171.95 & 0.93 & ${<}$0.01 & 3.53 \\ 
    json & 485.70 & 410.70 & 320.05 & 148.85 & 0.67 & 0.07 & 1.52 \\ 
    libxml2$^*$ & ${>}$29328.75 & ${>}$21462.88 & ${>}$6301.00 & ${>}$6318.63 & 1.00 & ${<}$0.01 & 4.65 \\
    openssl (A) & 1736.40 & 223.50 & 210.15 & 219.25 & 0.95 & ${<}$0.01 & 8.26 \\ 
    openssl (B)$^*$ & ${>}$26589.50 & ${>}$21431.00 & ${>}$12750.00 & $-$ & 1.00 & ${<}$0.01 & 2.09 \\ 
    pcre2 & 7994.80 & 6438.60 & 3900.30 & 3090.95 & 0.94 & ${<}$0.01 & 2.05 \\ 
    \hline
    \multicolumn{6}{c}{} & \textbf{Avg:} & 3.68${\times}$ \\
    \end{tabular}
    
\end{table}

The results are shown in \autoref{tab:bug}.
Overall we see that \isan can expose the corresponding bugs faster than \asan and \fuzzan.
Interestingly, some bugs (\verb+json+, \verb+pcre2+) are exposed faster using the coarse-grained \bisan configuration rather than the fine-grained \isan configuration.
These benchmarks are either not memory errors (\verb+json+), or are overflows beyond the object padding (\verb+pcre2+), and therefore benefit more from higher throughput.
Other bugs can only be detected by \isan and not \bisan.
For example, consider the \verb+openssl+ (B) bug (CVE-2017-3735), which occurs in the following function:

\lstset{language=C,
                basicstyle=\ttfamily\fontseries{m}\selectfont,
                deletekeywords={int},
                keywordstyle=\color{keyword}\ttfamily\fontseries{b}\selectfont,
                keywordstyle=[2]\color{type}\ttfamily\fontseries{b}\selectfont,
                keywordstyle=[3]\color{typecheck}\ttfamily\fontseries{b}\selectfont,
                keywordstyle=[4]\color{const}\ttfamily\fontseries{b}\selectfont,
                stringstyle=\color{red}\ttfamily,
                commentstyle=\color{comment}\ttfamily\fontseries{b}\selectfont,
                morekeywords=[2]{void, uint64_t,Token},
                morekeywords=[4]{NULL}
}
{\footnotesize
\begin{lstlisting}[escapechar=@]
unsigned X509v3_addr_get_afi(IPAddressFamily *f) {
    return (f != NULL && ...
           ? (f->addressFamily->data[0] << 8) |
              f->addressFamily->data[1] : 0);
}
\end{lstlisting}
}

\noindent
At runtime, it is allowable for \verb+f->addressFamily->data+ to point to a single byte array, meaning that accessing index \verb+1+ results in a single-byte overflow.
Since this (small) overflow only affects allocation padding, it is undetectable by the coarse-grained \bisan and is never exposed.
This example demonstrates how fine-grained \isan can detect more bugs in practice, and we therefore recommend \isan as the default configuration.

\result{\isan exposes bugs around 3.68${\times}$ faster than \asan.
\isan can also detect more bugs than \bisan in practice.
}

\myparagraph{RQ.5 Flexibility}
For this research question, we run additional experiments to demonstrate the overall flexibility of \isan with respect to fuzzer support, fuzzing modes, and scalability.
These results are intended to augment the main results of \autoref{tab:speed}.

\paragraph{Fuzzer Support.}
We chose AFL for our main benchmarks since it is natively supported by both \asan and \fuzzan.
However, \isan is not intended to be limited to one specific fuzzer or fuzzer version.
To demonstrate this, we have also integrated \isan into AFL++\cite{aflplusplus}, which is a more
modern fuzzer derived from standard AFL.
We conduct a set of limited experiments with
AFL++ against \asan, \isan, \bisan and Native (\fuzzan does not provide AFL++ support) and 6 subjects (\texttt{file}, \texttt{nm}, \texttt{objdump} \texttt{libpng}, \texttt{openssl}, \texttt{sqlite3}) from our main benchmarks (Table~\ref{tab:speed}).
We select subjects that (1) are included in FuzzBench framework \footnote{\url{https://github.com/google/fuzzbench}}, and (2) have a fuzzing harness that executes the similar program logic in both persistent (in-memory) and fork fuzzing modes.

\begin{table}[htbp]
\setlength{\tabcolsep}{4pt}
\centering\footnotesize
\caption{Average throughput (execs/sec) of AFL++.}
\label{tab:aflpp}
\begin{tabular}{r|rrrr|rr}
 \multirow{2}{*}{\textbf{Subject}} & \multirow{2}{*}{\textbf{\asan}} & \multirow{2}{*}{\textbf{\isan}}  & \multirow{2}{*}{\textbf{\bisan}} & \multirow{2}{*}{\textbf{Native}} & \multicolumn{2}{c}{\textbf{vs. \asan}} \\
  & & & & & \textbf{\isan} & \textbf{\bisan} \\ \hline
file	   & 284.71	& 431.69	& 487.774	& 725.87 & 51.62\% &	71.32\% \\
nm	    & 374.33	& 907.31	& 922.43	& 1072.73	& 142.38\%	& 146.42\% \\
objdump	& 375.50	& 897.62	& 998.35	& 1061.97	& 97.79\%	& 108.29\% \\
libpng	   & 929.07	& 1915.70	& 2038.84	& 2223.96 & 106.20\%	 &119.45\% \\
openssl	   & 191.95	& 212.48	& 238.054	& 395.75 & 10.70\% &	24.02\% \\
sqlite3	   & 128.48	& 398.28	& 412.974	& 500.90 & 209.99\%	 &221.43\% \\ 
\hline
\multicolumn{4}{c}{} & \multicolumn{1}{r}{\textbf{Avg:}} & 103.12\% & 115.15\%\\
\end{tabular}
\end{table}

The \autoref{tab:aflpp} shows the average throughput of AFL++ for both sanitizer and native execution.
We note that AFL++ implements several optimizations and improvements over standard AFL~\cite{aflplusplus}, and generally achieves an overall higher throughput.
Nevertheless, \isan achieves an overall improvement of 103.12\% versus \asan which is consistent with our main results.
The results also show that the performance of \isan is not tied to one specific fuzzer,
and that \isan can be integrated into other fuzzers.

\paragraph{Persistent (In-Memory) Mode Fuzzing.}
In the interest of completeness, we also tested \isan against (a.k.a., \emph{in-memory}) mode fuzzing.
Persistent mode aims to eliminate (or significantly reduce) the reliance on \verb+fork()+ to reset the program state between tests.
To do so, persistent mode relies on a developer-provided \emph{test harness} to manually reset the program state.
Unlike fork-mode fuzzing, persistent-mode is not automatic, so is not the default in standard fuzzing tools such as AFL and AFL++.

For this experiment, we use the test harnesses provided by FuzzBench.
Overall we measured a modest reduction in performance of -19.31\% for \isan and -11.47\% for \bisan compared to \asan using the same subjects as \autoref{tab:aflpp}.
Since \isan is specifically optimized for fork-mode fuzzing and uses a more complex instrumentation (see \autoref{fig:check}), a reduction in performance for persistent mode fuzzing is the expected result.
Nevertheless, the results show that (1) \isan can be applied to both fork and persistent modes, and (2) the
asymptotic performance of \isan over longer runtimes will eventually approach that of \asan.
These results also show that \fuzzan-style metadata switching may only have a marginal benefit under our sanitizer design.

\begin{table}[htbp]
\setlength{\tabcolsep}{2.3pt}
\centering\footnotesize
\caption{Average throughput (execs/sec) of AFL with sanitizers and native against Firefox (fork-mode).} 
\label{tab:firefox}
\begin{tabular}{r|rrrr|rr}
 \multirow{2}{*}{\textbf{Target}} & \multirow{2}{*}{\textbf{\asan}} & \multirow{2}{*}{\textbf{\isan}}  & \multirow{2}{*}{\textbf{\bisan}} & \multirow{2}{*}{\textbf{Native}} & \multicolumn{2}{c}{\textbf{vs. \asan}} \\
  & & & & & \textbf{\isan} & \textbf{\bisan} \\ \hline
ContentParentIPC	& 0.73	& 1.55	& 1.57	&3.02	&112.60\%	& 114.86\% \\
StunParser	        & 0.73	& 1.56	& 1.58	&3.04	&113.31\%	& 116.43\% \\
\hline
\multicolumn{4}{c}{} & \multicolumn{1}{r}{\textbf{Avg:}} & 112.95\% & 115.64\%\\
\end{tabular}
\end{table}

\paragraph{Scalability.}
To test the \emph{scalability} of \isan we test the Firefox browser version 91.3.0 (extended support).
To do so, we integrate \isan into the Firefox-customized version of AFL.
Since the Firefox project only supports fuzzing individual components (rather than whole program fuzzing)
so we select two targets (namely, \verb+ContentParentIPC+ and \verb+StunParser+) from Firefox-specific WebRTC/IPC subsystems respectively.
For these experiments we use fork-mode fuzzing.

The results are shown in \autoref{tab:firefox}.
Given the size of Firefox, the overall fuzzing throughput is much slower compared to the other benchmarks.
Nevertheless, both \isan and \bisan still outperform \asan with 112.60\% and 114.86\% improvement respectively.
These results are consistent with the other benchmarks, and demonstrate that \isan is scalable.

\myparagraph{RQ.6 False Detections}
The \isan sanitizer design allows for a small chance of false detections.
Our experiments comprise more than 19200 hours (${\sim}2.2$ years) of combined CPU time, during which no false detection was observed. 
This result is in line with expectations, where decades of CPU time would be required before we expect to observe the first false detection.

\result{No false detections were observed for \isan and \bisan during 19200 hours (${\sim}2.2$ years) of CPU time.}

\section{Related Work}\label{sec:relate}

We briefly summarize the related work in this section.

\myparagraph{Memory Poisoning}
As described in Section~\ref{sec:sanitizer},
one common approach (and also used by \isan) is to \emph{poison} memory that should not be accessed.
This approach has been implemented by many tools~\cite{asan, seward2005using, hasabnis2012light, bruening2011practical, sinha18rest, nethercote2007valgrind, fuzzan}.
Most existing tools implement memory poisoning using disjoint metadata, as opposed to the RET-based design of \isan.
However, as shown by our experiments, disjoint metadata can lead to high performance overheads under fuzzing environments.

\myparagraph{Guard Pages}
Another approach is to insert inaccessible guard pages between memory objects~\cite{bruce1993electri,micro2000pageheap}. Accessing a guard page will trigger a memory fault (\verb+SIGSEGV+) and terminate the program.
Both \emph{efence}~\cite{bruce1993electri} and GWP-ASAN~\cite{gwp-asan} implement this approach.
However, guard pages also mean that each protected object must reside in its own page of virtual memory---possibly leading to very high memory overheads.

\myparagraph{Canaries}
The \isan design has some similarities with \emph{stack canaries}~\cite{wagle2004stackguard}.
Stack canaries aim to detect stack-buffer-overwrites by delimiting stack buffers with a randomized canary value.
The idea has also been generalized to the heap~\cite{dmalloc}.
However, this approach is not instrumentation-based, and is limited to stack/heap-overwrites only. 
In contrast, \isan can detect more classes of memory error, including overreads and use-after-free.

\myparagraph{Low Fat Pointers and Pointer Tagging}
Another idea is to encode metadata within the pointer representation itself, such as with \emph{low fat pointers}~\cite{duck16heap, duck17stack}.
Unlike \isan, this is limited to object bounds errors only, and is not byte-accurate.
HWAsan~\cite{serebryany2018memory} also tags each pointer with a random value that is associated with a given object.
However, this approach still uses a disjoint meta data, and assumes a compatible instruction set architecture. 

\myparagraph{Probabilistic Methods}
DieHarder~\cite{gene10dieharder} randomizes the heap object layout in order to mitigate against
memory errors.
This approach is probabilistic, detects heaps errors only, and is primarily intended for hardening rather than bug detection via fuzz testing.
In contrast, \isan is designed to protect heap/stack/global objects with minimal fuzzing overheads.

\myparagraph{Improving Sanitizer Performance}
ASAP~\cite{wagner2015high} removes checks in code that is more often executed.
PartiSan~\cite{lettner2018partisan} creates different versions of the target program, in which some are more sanitized while others are not. 
These approaches essentially trade error coverage for improved performance.
SANRAZOR~\cite{zhang2021sanrazor} and ASAN\verb+--+~\cite{zhangdebloating} aim to only remove redundant checks, thereby preserving coverage.
Such optimizations are orthogonal to \isan and may be integrated as future work.

\myparagraph{Improving Fuzzing Performance}
FuZZan~\cite{fuzzan} similarly aims to optimize the combination of fuzz testing with sanitizers.
Unlike \isan, FuZZan still uses disjoint metadata, but may use more compact representation (based on \emph{RB-trees}) to minimize startup/teardown costs.
This approach can be more general, since different kinds of metadata can be supported, whereas \isan is specialized to the same class of memory errors that are covered by \asan.
Under our experiments, the disjoint metadata-free design of \isan further improves the throughput of fork-mode fuzzing.

\section{Discussion}

It is natural to combine fuzz testing with state-of-the-art memory error sanitizers, such as AddressSanitizer (\asan).
However, for fork-mode fuzzing, it has been shown that such a combination leads to poor performance~\cite{fuzzan}.
The underlying cause relates to the heavyweight metadata structures maintained by the sanitizer, leading to high program startup and teardown costs.

In this paper we introduced \isan---a lightweight sanitizer design based on \emph{Randomized Embedded Tokens} (RETs).
The basic idea is to poison memory directly using a randomized nonce, rather than relying on disjoint metadata.
We also show that the basic RET design can be further refined with an encoding of object boundary information---allowing for byte-accurate memory error detection without a significant difference in performance.
By eliminating the use of disjoint metadata, we show that \isan achieves a modest performance overhead of $1.27{\times}$ (or $1.14{\times}$ for coarse-grained checking) under fuzz testing environments, compared to a $2.36{\times}$ overhead for the state-of-the-art (\asan).
Our result helps to remove one of the main impediments to the fuzzer+sanitizer combination, and
enhances the overall effectiveness of memory error detection during fuzz campaigns.

\begin{acks}
We thank the anonymous reviewers for their insightful suggestions. This research was partially supported by the National Research Foundation Singapore (National Satellite of Excellence in Trustworthy Software Systems).
\end{acks}

\balance
\newpage
\bibliographystyle{ACM-Reference-Format}
\bibliography{canary}

\end{document}